**Gallium Oxide Heterojunction Diodes for Improved High-Temperature Performance**

Shahadat H. Sohel*, Ramchandra Kotecha, Imran S Khan, Karen N. Heinselman, Sreekant Narumanchi, M Brooks Tellekamp*, Andriy Zakutayev*

*National Renewable Energy Laboratory, 15013 Denver West Blvd., Golden 80401, CO, United States of America*

E-mail: shahadat.sohel@nrel.gov, brooks.tellekamp@nrel.gov, andriy.zakutayev@nrel.gov

**Abstract**

$\beta$-$Ga_2O_3$ based semiconductor devices are expected to have significantly improved high-power and high-temperature performance due to its ultra-wide bandgap of close to 5 eV. However, the high-temperature operation of these ultra-wide-bandgap devices is usually limited by the relatively low 1-2 eV built-in potential at the Schottky barrier with most high-work-function metals. Here, we report heterojunction p-NiO/n-$\beta$-$Ga_2O_3$ diodes fabrication and optimization for high-temperature device applications, demonstrating a current rectification ratio ($I_{ON}/I_{OFF}$) of more than $10^6$ at 410°C. The NiO heterojunction diode can achieve higher turn-on ($V_{ON}$) voltage and lower reverse leakage current compared to the Ni-based Schottky diode fabricated on the same single crystal $\beta$-$Ga_2O_3$ substrate, despite charge transport dominated by interfacial recombination. Electrical characterization and device modeling show that these advantages are due to a higher built-in potential and additional band offset. These results suggest that heterojunction p-n diodes based on $\beta$-$Ga_2O_3$ can significantly improve high-temperature electronic device and sensor performance.

# 1. Introduction

$\beta$-Ga$_2$O$_3$ has recently gained significant interest for high-power electronic applications[1] due to the availability of low-cost native substrates[2], an ultra-wide bandgap[3], and facile n-type doping control[4]. Excellent unipolar device performance has been already reported using different edge termination techniques like ion-implantation termination, field-plate terminations, and trench structure with field plates[5-8]. A breakdown voltage of 2.89 kV has recently been achieved[8] using vertical trench Schottky diodes, which leads to a Baliga figure of merit (BFOM)[9] of 0.8 GW/cm$^2$.

The ultra-wide bandgap of $\beta$-Ga$_2$O$_3$ also allows it to have significantly improved high-temperature performance due to ultra-low intrinsic carrier concentration. This will allow $\beta$-Ga$_2$O$_3$ power devices and sensors to have much higher operating temperature range compared to existing Si-based devices, which are limited to a maximum operating temperature of 150°C[10]. High-temperature operation of power devices and sensors will allow for real-time monitoring of a number of high-temperature processes applicable to power plants, coal mines, gas turbines and aerospace systems[11]. This will increase the system reliability and increase the efficiency by reducing failures of critical components [12].

Despite significant progress in unipolar device performances for power applications, the complete utilization of the $\beta$-Ga$_2$O$_3$ material potential for high-temperature applications is still limited by poor reverse bias performance, due to the relatively small Schottky barrier height when using ideal Ni or Pt metal Schottky contacts[13]. In addition, the complicated field engineering techniques add significant challenges to the device fabrication process. For further improvements to the reverse bias performance of $\beta$-Ga$_2$O$_3$ devices, bipolar device designs need to be explored. However, due to a lack of suitable p-type dopants and polaronic hole transport[14], a high-quality p-n homojunction is currently unachievable.

As an alternative solution to this problem, p-n heterojunctions using n-type $\beta$-Ga$_2$O$_3$ and other p-type semiconductors have been recently explored by several groups[13, 15-17]. Different p-type layers have been adopted to make contact to n-type $\beta$-Ga$_2$O$_3$. Among them, Cu$_2$O and NiO are some of most promising and chemically simple candidates, and high breakdown voltages have been demonstrated using these heterojunction p-n diodes. Breakdown voltages of 1.49 kV and 1.86 kV have been reported using Cu$_2$O/$\beta$-Ga$_2$O$_3$ and NiO/$\beta$-Ga$_2$O$_3$ p-n heterojunctions respectively[13, 16]. Among the p-type oxide materials, NiO has excellent potential for optoelectronics and power devices due to its transparency and high p-type conductivity[18]. Though p-n heterojunction diodes based on $\beta$-Ga$_2$O$_3$ have explored, there is little work on optimizing the heterojunction performance for high-temperature applications.

In this work, we demonstrate modeling, fabrication, and characterization of NiO/$\beta$-Ga$_2$O$_3$ heterojunctions for high-temperature applications. Textured polycrystalline p-NiO layers are grown using pulsed laser deposition (PLD) on single-crystal n-type $\beta$-Ga$_2$O$_3$ without any drift layer, patterned using a reactive ion etch (RIE) process, and subjected to a high-temperature annealing step. The resulting heterojunction devices show excellent high-temperature performance with a current rectification ratio ($I_{ON}/I_{OFF}$) of more than $10^6$ at 410 °C. These heterojunction metrics are much better compared to Ni-based Schottky contacts deposited directly on n-type $\beta$-Ga$_2$O$_3$ with the same doping characteristics. The optimized heterojunction p-NiO/n-$\beta$-Ga$_2$O$_3$ diode high-temperature performance is achieved due to larger built-in electric field and additional band offset, according to electrical characterization and device modeling reported in this paper. These results show the promise of $\beta$-Ga$_2$O$_3$ heterojunction diodes for high-temperature applications in sensors and other electronic devices.

## 2. Methods

All devices, both Schottky and heterojunction diodes, were fabricated on commercially available n-type Sn-doped (001) orientation $\beta$-Ga$_2$O$_3$ substrates grown by Novel Crystal Technology, Japan. No unintentionally doped (UID) drift layer was used for this experiment, which was aimed at optimizing the high-temperature performance of the $\beta$-Ga$_2$O$_3$-based devices rather than drift layers. The Ga$_2$O$_3$ samples were cleaned using organic solvents (Acetone, IPA, and DI for 5 minutes each) prior to each processing step. In both cases, a back Ohmic contact of Ti/Au (20 nm/100 nm) was deposited using an e-beam evaporator. The device was then annealed via Rapid Thermal Annealing (RTA) at 550 °C in N$_2$ for 2 minutes to improve the ohmic contact.

For the Ni Schottky diodes, circular pads of $\beta$-Ga$_2$O$_3$ with different diameters ranging from 50 μm – 300 μm were patterned using lithography with an ABM mask aligner and exposure system, and 100 nm Ni Schottky contacts were deposited using the e-beam evaporator. Figure 1(a) shows the schematic of the resulting baseline Ni/$\beta$-Ga$_2$O$_3$ Schottky diodes. A doping concentration of n = $4\times10^{18}$ cm$^{-3}$ was determined in $\beta$-Ga$_2$O$_3$ by *C-V* measurement as shown in Fig. 1(b).

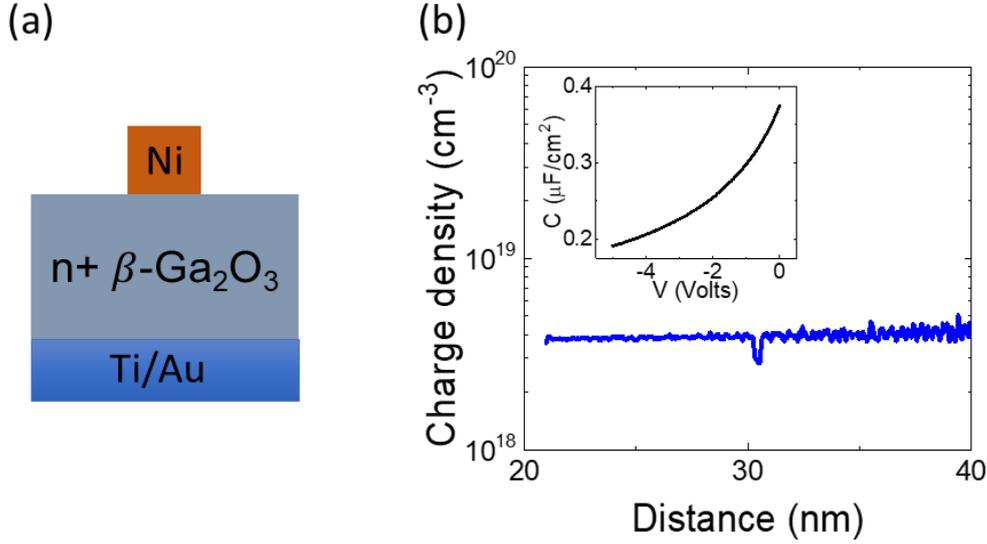

Fig. 1: Baseline Ni/β-Ga$_2$O$_3$ devices: (a) Schematic of vertical Ni/β-Ga$_2$O$_3$ Schottky diode and (b) charge distribution extracted from C-V measurements (inset) on β-Ga$_2$O$_3$ ($\epsilon_r$ = 12.4)[19] showing a constant n-type doping profile of 4×10$^{18}$ cm$^{-3}$ within the depletion width of approximately 40 nm.

For the heterojunction device fabrication, circular NiO top contacts of different diameters ranging 50 μm – 300 μm were patterned by photolithography, and a Ti/Au (20 nm/100 nm) metal stack was deposited on top of the NiO. To reduce current spreading, mesa etching with a separate mesa-etching mask with 5-10 μm spacing from Schottky contacts was performed using Ar-based ICP-RIE (Ar flow 30 sccm, pressure 5 mTorr, RIE power 100 W, ICP power 700 W). An etch rate of ~25 nm/minute was determined from the profilometer using NiO layers on glass substrates coloaded during the PLD growth. To further improve the NiO layer quality and reduce the leakage, a 10 minute anneal at 200°C was performed in air on a hotplate. Figure 2(a) shows the schematic of the completed heterojunction p-NiO/n-β-Ga$_2$O$_3$ diodes.

The NiO films were deposited on β-Ga$_2$O$_3$ substrates by PLD using a Coherent COMPexPro 205 KrF excimer laser operating at 248 nm and a pulse duration of 10 ns. The laser, with an energy of 350 mJ and a repetition rate of 10 Hz, was focused onto a rotating 1" diameter commercial NiO target. PLD conditions including O$_2$ pressure and substrate temperatures were optimized for low sheet resistance. X-ray diffraction (XRD) using Brooker Discover D8 system with 2D detector for the different PLD growth temperatures is shown in Fig. 2(b). The optimized condition that was used for the NiO deposition was 85 sccm $O_2$ flow resulting in 50 mTorr chamber pressure, and a substrate temperature of 350°C. The deposited films were polycrystalline with preferential NiO (111) and (200) orientations. For 50,000 laser pulses, a 200-nm-thick NiO layer was grown with a sheet resistivity of $3 \times 10^6$ $\Omega/sq$ as measured by 4-point probe.

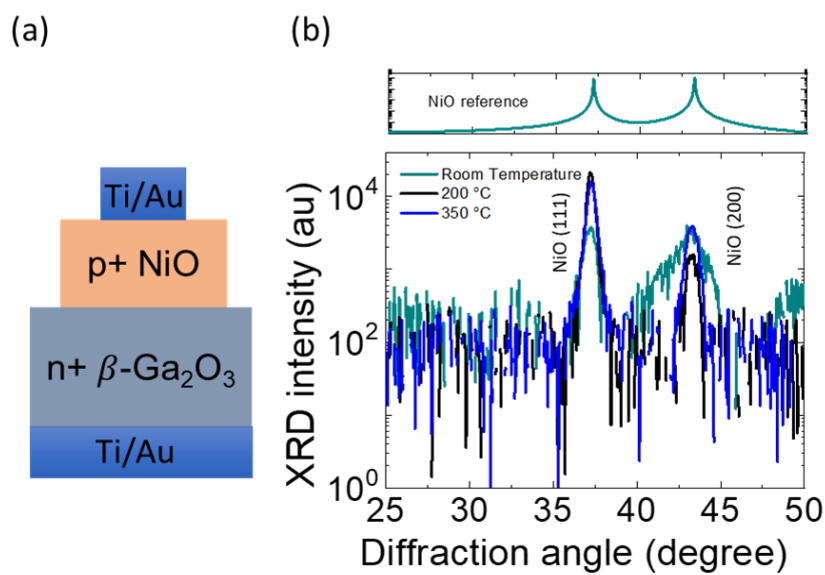

Fig. 2: Heterojunction p-NiO/n-$\beta$-$Ga_2O_3$ devices: (a) Schematic of vertical heterojunction p-NiO/n-$\beta$-$Ga_2O_3$ diodes and (b) XRD of 200-nm-thick NiO grown at various substrate temperatures by PLD using 50,000 pulses at 350 mJ in 50 mTorr $O_2$, showing polycrystalline NiO films with preferential orientation.

## 3. Results and discussions

We found that the mesa etching is critical to reduce current spreading in the p-NiO layer[20], demonstrated by decreased leakage current in Figure 3. Additionally, annealing at 200 °C for 10 min in air further improved the device performance, presumably by improving the interface quality. The mesa etch reduces the current spreading and hence, increases the forward bias current density. The overall result is higher turn-on voltage (measured voltage at 0.5 A/cm$^2$), and lower leakage current, critical to p-n heterojunction device performance. The minimum ideality factor decreases from 3 to 2 after annealing and mesa steps, which shows a possible improvement of the interface layers between NiO and $Ga_2O_3$. For all measurements contacts with 200 μm diameter are used. The measurement current compliance was set to 1 mA absolute current, which is equivalent to 3.2 mA/cm$^2$. Figure 3(b) shows the increase in leakage current with increasing reverse voltage. The heterojunction p-NiO/n-$\beta$-$Ga_2O_3$ diode shows a reverse leakage of 1 mA/cm$^2$ at a reverse bias of 15 V This leakage level is achieved without any drift layer. By implementing a UID drift layer between p and n layer, the leakage can be suppressed significantly, and a high breakdown voltage can be achieved.

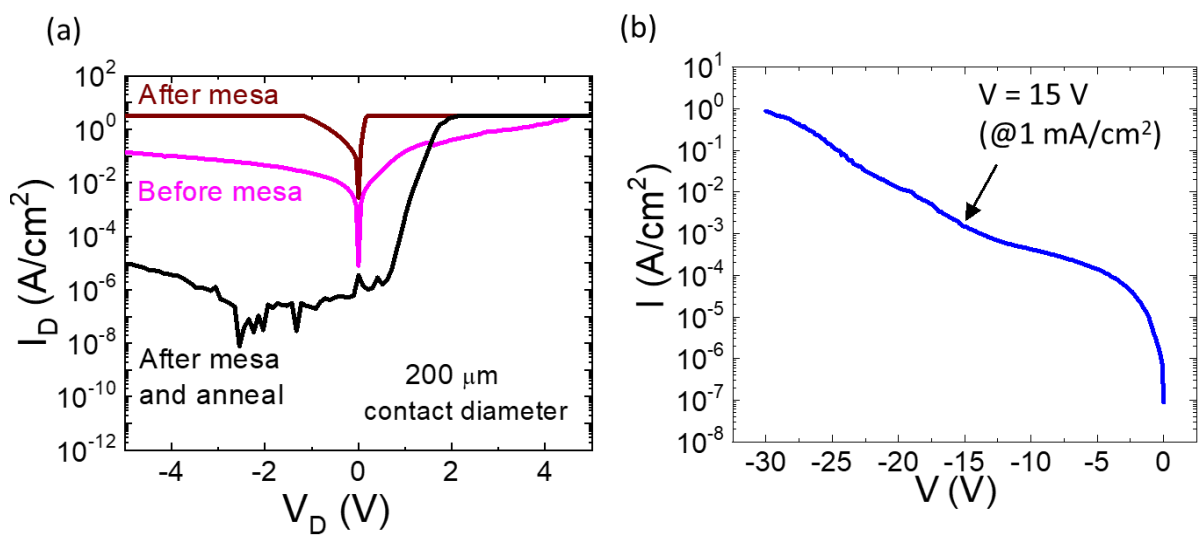

Fig. 3: (a) Effect of anneal and mesa etching on heterojunction p-NiO/n-$\beta$-$Ga_2O_3$ diodes. The mesa etch is effective at reducing current spreading and decreasing the leakage current. After annealing, the nearly ideal diode behavior is recovered through improvement of the interface and the p-NiO material quality. (b) Room temperature breakdown measurement after mesa and annealing shows a 15V breakdown (measured at 1 mA/cm$^2$) of the p-NiO/n-$\beta$-$Ga_2O_3$ diodes.

### 3.1 Room-temperature characterization

Figure 4 shows the forward *I-V* characteristics along with the extracted ideality factor for both Ni Schottky contact and NiO heterojunction p-contact to $\beta$-$Ga_2O_3$ at room temperature. The NiO heterojunction contact increases the turn-on voltage to ~1.6 V compared to ~1.3 V for the Ni Schottky contact. The minimum ideality factor for the Ni Schottky contact is 1.09, which indicates near ideal diffusion-dominated charge transport. In contrast, the minimum ideality factor for the heterojunction p-NiO/n-β-Ga$_2$O$_3$ diodes is 2, which suggests transport is dominated by trap-assisted recombination. This non-ideal behavior is likely due to interface traps at the heterointerface generated during PLD growth.

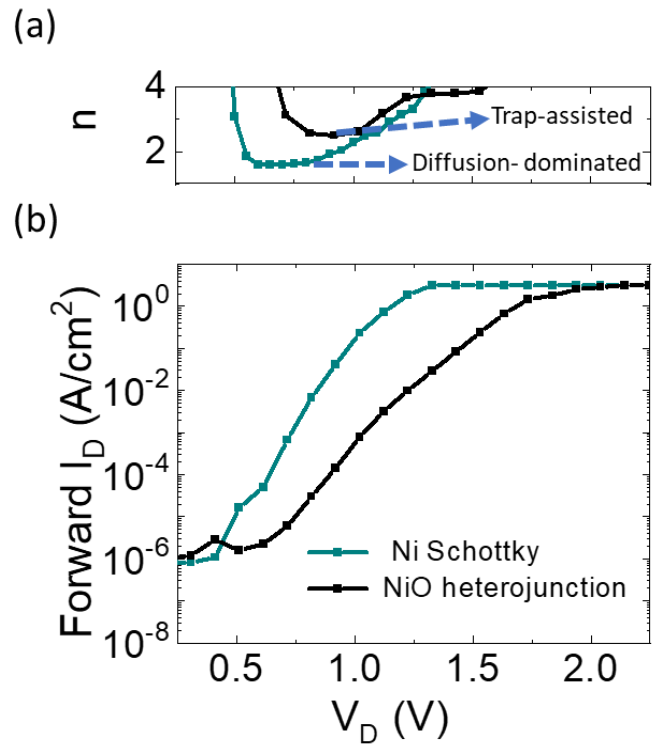

Fig. 4: Forward bias I-V characteristics and ideality factor for Ni/β-Ga$_2$O$_3$ Schottky and NiO/β-Ga$_2$O$_3$ p-n heterojunction diodes. The NiO heterojunction is effective at increasing the turn-on voltage, however there is a clear change from ideal diode diffusion-dominated transport in the Ni Schottky diode to trap-assisted recombination-dominated transport in the NiO heterojunction.

Figure 5 shows the Mott-Schottky analysis ($1/C^2$ versus V plot) for determining built-in potential, derived from C-V characteristics measured at a frequency of 100 kHz for both Ni/β-Ga$_2$O$_3$ Schottky and NiO/β-Ga$_2$O$_3$ p-n heterojunction diodes. The built-in potential for the Ni Schottky diode is ~1.3 V, which matches closely with the reported values in literature (1.22 eV)[21], and also matches the measured room-temperature turn-on voltage. For the NiO heterojunction, we observe an increased built-in potential of around 1.9 V. This value for the built-in potential varies slightly from the theoretical prediction (1.4 V)[16]. The variation could be attributed to non-ideal interface condition leading to potential Fermi level pinning for the heterojunction and polycrystalline crystal structure of NiO. The built-in potential for both the Schottky and the p-n diodes shown in Fig.5 match closely with the respective turn-on voltage shown in Fig.4. This increased built-in potential will act as a barrier to the carriers in reverse bias conditions and is expected to reduce the reverse bias leakage current.

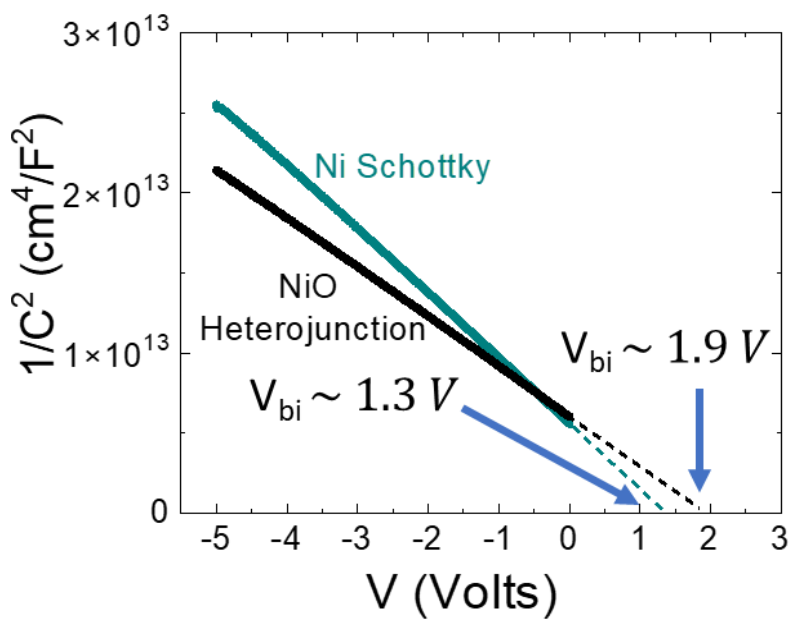

Fig. 5: Mott-Schottky analysis ($1/C^2$ versus V plot) measured at a frequency of 100 kHz for both Ni/$\beta$-Ga$_2$O$_3$ Schottky and NiO/$\beta$-Ga$_2$O$_3$ p-n heterojunction diodes shows improved built-in potential for the heterojunction diode compared to the Schottky diode.

*3.2 High-temperature performance*

To measure the high-temperature performance of the devices, we bonded the sample on ITO-coated glass using silver paint to maintain both thermal and electrical contact to the back of the sample. The devices were then measured using a custom-built high-temperature stage. The stage heater setpoint was varied from room temperature to 500°C, which resulted in a thermocouple-measured maximum sample temperature of 410 °C[22]. Figure 6 shows the high-temperature performance for both Ni/$\beta$-Ga$_2$O$_3$ Schottky diodes and p-NiO/n-$\beta$-Ga$_2$O$_3$ heterojunction diodes.

From Fig. 6(a), we can see that increasing temperature results in a decreasing turn-on voltage with a minimum V$_{on}$ of around 0.4 V at the highest 410 °C measurement temperature, and a significant increase in the reverse leakage (rectification ratio reduces to $10^3$ at 410 °C, compared to $10^6$ at room temperature) for the Ni/$\beta$-Ga$_2$O$_3$ Schottky barrier diode. This expected behavior is due to an increase in thermionic emission and thermionic field emission over the Schottky barrier[23].

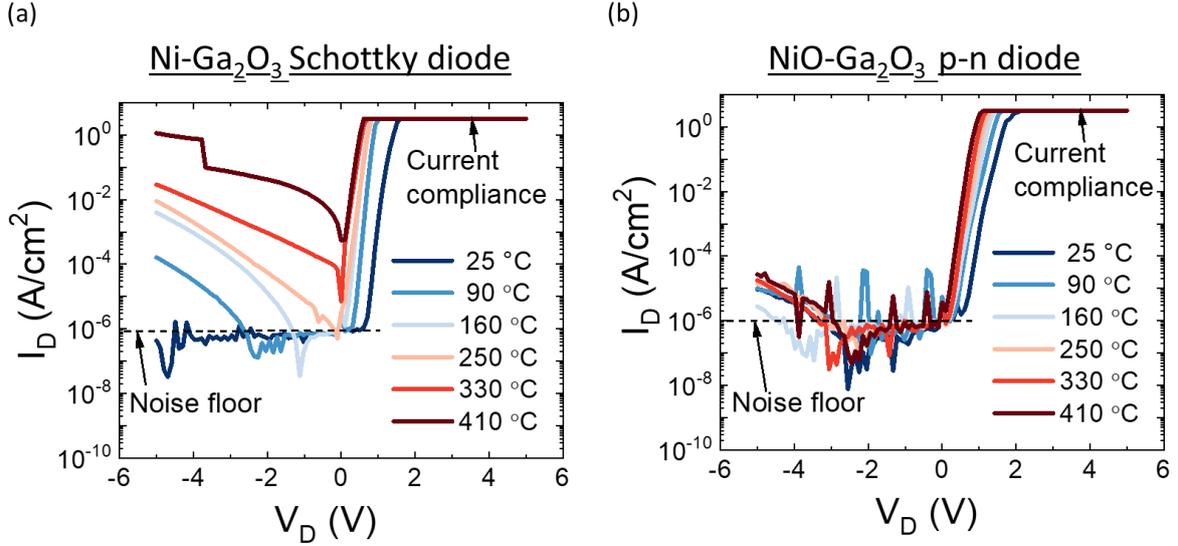

Fig. 6: High-temperature rectification characteristics for Ni/$\beta$-Ga$_2$O$_3$ Schottky and heterojunction p-NiO/n-$\beta$-Ga$_2$O$_3$ diodes show improved high-temperature behavior from the heterojunction diode. Measurements are done on 200 μm contact pads with 1 mA current compliance.

For the heterojunction p-NiO/n-$\beta$-Ga$_2$O$_3$ diodes, the $V_{on}$ at 410 °C is about 0.8 V, higher than for the Ni/$\beta$-Ga$_2$O$_3$ Schottky barrier diode. Reverse leakage current does not exhibit any meaningful change for the whole measurement range and the heterojunction shows a rectification ratio > $10^6$ even at 410 °C. This improved high-temperature behavior can be attributed to improved built-in potential for the NiO/$\beta$-Ga$_2$O$_3$ p-n heterojunction diode compared to Ni/$\beta$-Ga$_2$O$_3$ Schottky diode, and the behavior can be further improved by improving the interface quality. This result shows better performance compared to the previous reports of $\beta$-Ga$_2$O$_3$-based high-temperature devices, which showed a maximum operating temperature of 350 °C[24]. However, there is still room for improvement, because other ultrawide-bandgap materials like SiC have been shown to work up to 600 °C using p-i-n configuration[12]. For a direct comparison, Ni/4H-SiC Schottky diodes show an on-off ratio of $10^6$ up to 300 °C[25] and AlN/4H-SiC heterojunction diode shows an on-off ratio of ~$10^5$ up at 450K (177 °C)[26].

As a summary of these experimental results, Figure 7(a) shows the change in the turn-on voltage (extracted at 0.5 A/cm$^2$) for both the Schottky and the heterojunction diode, and Fig. 7(b) shows the change in the reverse leakage for the Schottky contact extracted at -3 V bias. Heterojunction reverse leakage was not plotted as the values were near the noise floor of the equipment and hence, did not show any meaningful trend. Further optimization resulting in better heterojunction interface and NiO growth could result in further improved device performance from heterojunction p-NiO/n-$\beta$-Ga$_2$O$_3$ diodes.

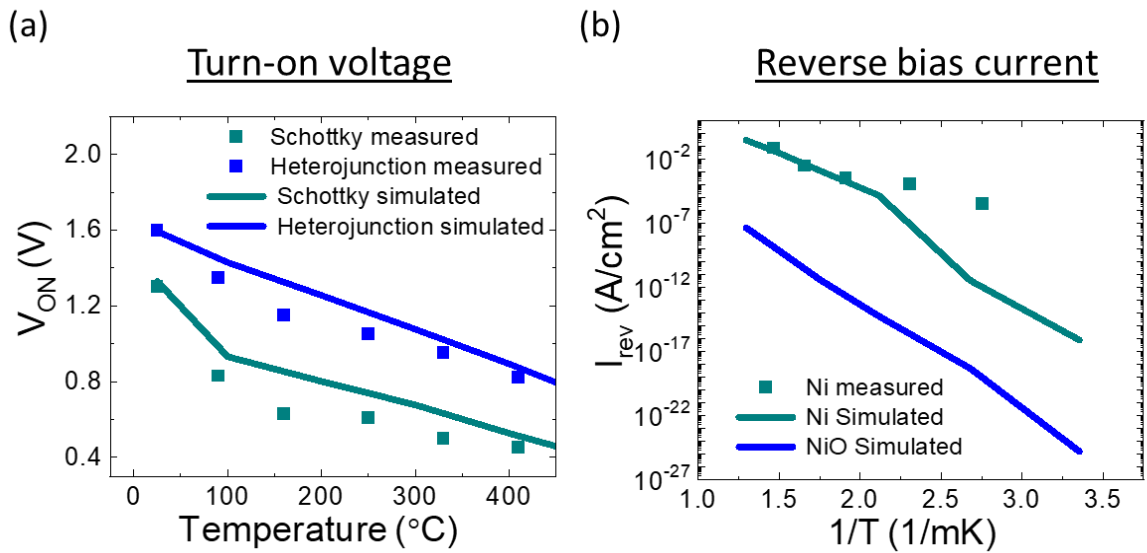

Fig. 7: Measured and simulated (a) Turn-on voltage for both Ni/$\beta$-Ga$_2$O$_3$ Schottky and NiO/$\beta$-Ga$_2$O$_3$ p-n heterojunction diodes, and (b) reverse leakage current for Ni/$\beta$-Ga$_2$O$_3$ Schottky diode shows higher turn-on voltage and lower reverse leakage current for the heterojunction diode.

### *3.3 Modeling of Ga$_2$O$_3$ diodes*

To analyze and compare the performance of the vertical Schottky diode and heterojunction diode, technology computer aided design (TCAD) models were created using Sentaurus TCAD software[27], as described in our previous publication[28]. The device structures were developed using the device structures and doping information as described in Figs. 1 and 2. The Schottky diode is modeled using thermionic emission and tunneling charge transport mechanisms. For the heterojunction model, an additional temperature-dependent trap-assisted tunneling is used because of the experimental results shown in Figs. 3 and 4. The Ni contact is used as the Schottky contact for the diode, since Ni showed better interface quality and closer-to-ideal diode behavior compared to Pt in our previous experiments[22], likely due to the effect of water trapped at the interface[29]. The metal work function for Ni is taken to be 5.05 eV for the simulations[30]. The back contact was modeled as an Ohmic contact. Table I shows the parameters used for NiO and $\beta$-Ga$_2$O$_3$ for the simulations[28, 31, 32].

TABLE I

MATERIAL PARAMETERS USED FOR SIMULATIONS

| Properties | Ga$_2$O$_3$ | NiO |
|---|---|---|
| Bandgap (eV) | 5.02 | 3.6 |
| Electron Affinity (eV) | 3.618 | 1.45 |
| Relative dielectric constant ($\varepsilon_r$) | 10 | 11.75 |
| Electron mobility (cm$^2$/Vs) | 100 | 50 |
| Lattice Heat Capacity (J/ K.cm$^3$) | 3.332 | 1.67 |
| Thermal Conductivity (W/cm.K) | 0.27 | 0.33 |
| Electron Saturation Velocity (cm/s) | $2\times10^7$ | - |
| Electron Effective mass | 0.28m$_0$ | 0.121m$_0$ |

Fig. 8(a) shows the energy-band diagram for the Schottky diodes at room temperature as well as 410 °C, and Fig. 8(b) shows the simulated band diagram for the heterojunction p-NiO/n-$\beta$-Ga$_2$O$_3$ diodes. These band diagrams, using the electron affinity model, show that the NiO-Ga$_2$O$_3$ junction forms a type-II heterojunction with staggered band alignment used for the modeling. The simulated band diagram shows that the heterojunction band diagram has a higher barrier for the carriers due to the p-type contribution and the type-II band alignment. This helps to suppress reverse leakage current. At higher temperatures, the bandgap for the semiconductor slightly decreases, but the barrier is still significantly higher for the heterojunction diode compared to the Schottky diode.

Figure 7 shows a comparison of simulated and measured temperature-dependent behavior of the Schottky diode and the heterojunction diode. The simulated turn-on voltage of the heterojunction is significantly higher than Schottky diode for the entire temperature range, in agreement with experimental results. The reverse current density of the heterojunction diode is significantly lower than the Schottky diode in direct agreement with the experimentally measured results. The leakage current in the heterojunction diode is modeled to increase with temperature due to temperature-dependent trap-assisted tunneling current, but even at the highest measurement temperature, the experimental current is still below the equipment measurement threshold.

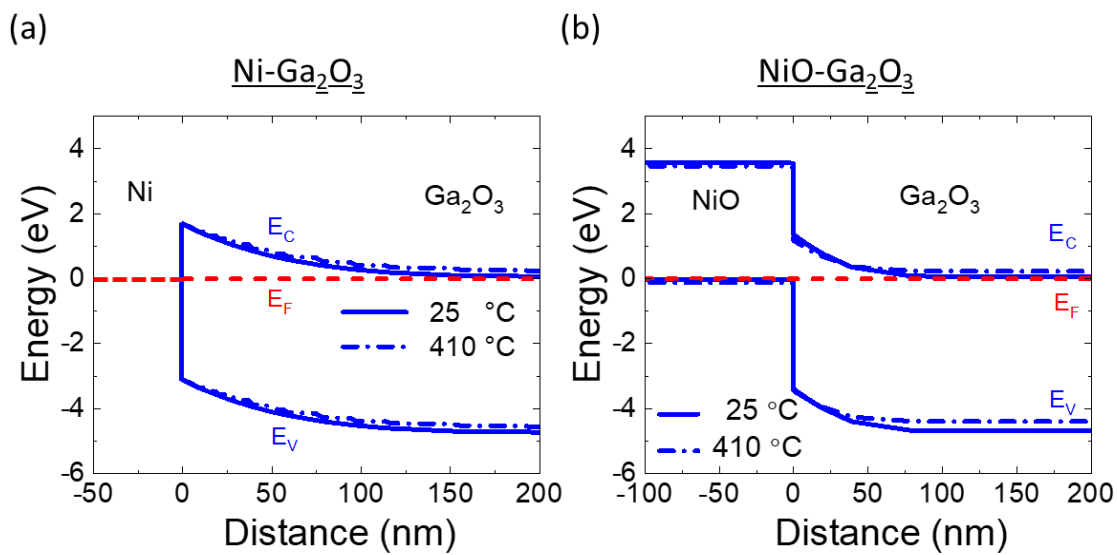

Fig. 8: Simulated energy-band diagram of (a) Ni/$\beta$-Ga$_2$O$_3$ Schottky diode and (b) heterojunction p-NiO/n-$\beta$-Ga$_2$O$_3$ diode at 25°C (solid) and 410°C (dot-dashed) shows higher barrier for the heterojunction at both room temperature and at higher temperature.

Overall, the trends for the change in turn-on voltage and leakage current shown in Fig.7 match well with the experimental results, but there are a few notable differences. In particular, the turn-on voltage for both the Schottky and heterojunction diodes are slightly lower than the simulated values. This difference is likely due to the interface traps and the polycrystalline structure of the NiO layer (for the heterojunction). In addition, the leakage current measured for the Ni Schottky diverges at lower temperatures, which may be due to hitting the noise floor on the measurement equipment. Despite these small differences, it can be concluded that heterojunction diodes offer significant improvement in performance compared to Schottky diode, and they mitigate the current limitations of p-type doping in $\beta$-Ga$_2$O$_3$.

## 4. Conclusion

High-temperature operation of $\beta$-Ga$_2$O$_3$ diodes is demonstrated using a p-n device structure with p-NiO heterojunction contact. The PLD NiO growth as well as etch and annealing device fabrication steps resulted in improved high-temperature performance compared to Ni /$\beta$-Ga$_2$O$_3$ Schottky diodes. A diode on-off ratio of more than $10^6$ is achieved up to 410°C for the p-n heterojunction diode. The improved performance is attributed to the increased built-in potential and additional band offset of the p-n heterojunction diode. The temperature-dependent turn-on voltage of the p-n heterojunction diode is shown to deviate from ideal behavior, decreasing with temperature more significantly than the model, which is attributed to the presence of traps at a non-ideal crystalline interface. The interface traps can be also deduced from fitting the measured J-V curves to the diode equation model, with extracted ideality factor that indicates trap-assisted current conduction as the primary mechanism, rather than ideal diffusion behavior. These results indicate that the heterojunction p-contact-based $\beta$-Ga$_2$O$_3$ can lead to high- performing devices suitable for extreme high-temperature operation.

**Acknowledgments**

This work was authored by the National Renewable Energy Laboratory (NREL), operated by Alliance for Sustainable Energy, LLC, for the U.S. Department of Energy (DOE) under Contract No. DE-AC36-08GO28308. Funding provided by the Office of Energy Efficiency and Renewable Energy (EERE) Advanced Manufacturing Office. A portion of the research was performed using computational resources sponsored by the U.S. Department of Energy's Office of Energy Efficiency and Renewable Energy and located at the National Renewable Energy Laboratory.